\documentclass[aip,amsmath,amssymb,reprint,jcp]{revtex4-1}  
\usepackage[T1]{fontenc}
\usepackage{setspace}
\usepackage{graphicx}
\usepackage{textcomp}
\usepackage{array,booktabs}
\usepackage{bm}
\usepackage{color}

\begin{document}

\title{Reverse Monte Carlo modeling of liquid water with the explicit use of the SPC/E interatomic potential} 

\author{Ildik\'o Pethes}
\email[]{pethes.ildiko@wigner.mta.hu}
\author{L\'aszl\'o Pusztai}

\affiliation{Institute for Solid State Physics and Optics, Wigner Research Centre for Physics, Hungarian Academy of Sciences, Konkoly-Thege M. \'ut 29-33, 1121, Budapest, Hungary}

\date{\today}

\begin{abstract}
Reverse Monte Carlo modeling of liquid water, based on one neutron and one X-ray diffraction data set, applying also the most popular interatomic potential for water, SPC/E, has been performed. The strictly rigid geometry of SPC/E water molecules had to be loosened somewhat, in order to be able to produce a good fit to both sets of experimental data. In the final particle configurations, regularly shaped water molecules and straight hydrogen bonding angles were found to be consistent with diffraction results. It has been demonstrated that explicit use of interatomic potentials in RMC has a role to play in future structural modeling of water and aqueous solutions. 
\end{abstract}

\maketitle

\section{Introduction}

Water, the most common liquid on Earth, has been (one of) the most frequently investigated substance for thousands of years. The microscopic structure of liquid water, which is a fundamental piece of information for researchers in many fields of research, is one of the oldest not fully resolved problems. Despite the dozens of publications appearing year after year, our knowledge is still uncertain in this respect \footnote{see e. g. http://www1.lsbu.ac.uk/water/.}.
The unique behavior of water stems from the hydrogen bonded network of the molecules. However, crucial features of this network, the average number of the hydrogen bonds per molecules, or the intermolecular O-H bond distance are still somewhat controversial\cite{WernetScience2004, Head-Gordon2006, Leetmaa2008, Pusztai1999}.

The microscopic structure of liquid water has been studied by different spectroscopic and scattering techniques, for example small and wide angle X-ray scattering \cite{Skinner2013, Hura2003, Narten1971, Huang2009, Clark2010PNAS}, neutron diffraction \cite{Soper2015, Zeidler-Salmon2012, Narten1982}, and X-ray absorption and emission spectroscopies \cite{WernetScience2004, Myneni2002, Smith2004,  Tokushima2008, Fuchs2008}. X-ray diffraction is suitable for the determination of oxygen-related correlations, but it is less sensitive to hydrogen. Neutron diffraction with H/D isotopic substitution can be useful for the detection of hydrogen-hydrogen and hydrogen-oxygen correlations, since H has negative coherent scattering length, $b_c^{\mathrm{H}}$ = -3.7406 fm, while  $b_c^{\mathrm{D}}$=6.671 fm. However, the determination of the coherent structure factor from the measured neutron scattering intensities is difficult due to the large incoherent scattering cross section of H, and the strong inelasticity effects caused by the similar masses of H nuclei and incident neutrons. There are numerous attempts known for resolving these issues, e.g. via oxygen isotope substitution \cite{Zeidler-Salmon2011} and polarized neutron diffraction with polarization analysis \cite{Temleitner2015}.

Various computer simulation methods have also been applied in order to gain real-space correlation functions: the story started with  Monte Carlo simulations nearly fifty years ago \cite{Barker1969} and continued with molecular dynamics (MD) (for a review, see e.g. Ref. \onlinecite{Guillot2002}), ab initio MD (see e.g. Ref. \onlinecite{Szalewicz2009}) and Reverse Monte Carlo (RMC) (see e.g. Ref. \onlinecite{Pusztai1999}) simulations. Monte Carlo and molecular dynamics simulations are based on the intra- and intermolecular interactions between particles, thus their outcome is determined by the chosen force fields. During the past decades, several water models with different force field parameters have been developed  that have been fitted to some chosen experimental data (for reviews, see \cite{Hura2003, Guillot2002, Jorgensen2005}).

In Reverse Monte Carlo structural modeling \cite{McGreevy1988} large 3-dimensional particle configurations are generated that are consistent with all the supplied input data sets within their uncertainty. Any experimental (and/or theoretical quantities) that can be expressed in terms of the atomic coordinates may be fitted simultaneously. Conventional RMC algorithms \cite{McGreevy1988,Tucker2001,Evrard2005,Gereben2007} are not able to take energetic considerations into account, although the so-called 'hybrid RMC' scheme by Opletal et al. \cite{Opletal2002,Opletal2008} does operate with specific potential parameters. Another method of structural modeling, the 'Empirical Potential Structure Refinement' (EPSR) \cite{Soper1996,Soper2005}, starts with known interatomic potential parameters that are varied during the calculations. EPSR has been applied to liquid water over the past 20 years, starting with the original publication in 1996 \cite{Soper1996}, to a very recent extensive paper \cite{Soper2015}.

The RMC\_POT algorithm (and the software that makes use of it) \cite{Gereben2012} combines traditional RMC modeling with some features of standard molecular simulations. Instead of 'Fixed Neighbour Constraints' (FNC) used previously (see, e.g., Ref. \onlinecite{Evrard2005}), the RMC\_POT algorithm keeps molecules together via (more or less) flexible intramolecular forces: bond stretching, angle bending and dihedral stretching potential functions (see, e.g., Ref. \onlinecite{GROMACS}). Besides these, RMC\_POT can handle intermolecular potentials of arbitrary complexity: Coulomb and Lennard-Jones energies can be easily calculated. It is worth stressing that while in EPSR \cite{Soper1996} potential parameters are being modified continuously, RMC\_POT keeps all intermolecular terms intact.

Total scattering structure factors (TSSF) obtained from neutron and X-ray diffraction experiments were studied by the conventional RMC technique, using FNC, in an early publication \cite{Pusztai1999}. The RMC technique has also been used to investigate the compatibility of structure factors and partial pair correlation functions (PPCF) obtained by different methods in Refs. \onlinecite{Steinczinger2012, Pethes2015}. The conventional RMC technique was used to compare different water potential models via PPCFs obtained by MD simulations and structure factors from neutron and X-ray diffraction measurements in Refs. \onlinecite{Pusztai2008,Steinczinger2013}.

In this study the suitability of the RMC\_POT algorithm for the determination of the structure of water is explored. X-ray and neutron  diffraction structure factors from Refs. \onlinecite{Skinner2013,Narten1971,Soper1997} were applied here as input data. Of the many possibilities, the most frequently used SPC/E ('Extended Simple Point Charge') water potential model \cite{Berendsen1987} was selected for the purpose. We are aware that the intramolecular O-H distance of the SPC/E model is very slightly (by about 0.02 \AA ) longer than suggested by most neutron diffraction experiments (see, e.g., Ref. \onlinecite{Zeidler-Salmon2012}); however, the overwhelming popularity of SPC/E over the past decades justifies the choice of this potential for the first RMC\_POT study on liquid water. For generating starting particle arrangements for RMC\_POT, as well as for reference purposes, molecular dynamics computer simulations with the SPC/E water potential \cite{Berendsen1987} have been carried out as an initial step.

\section{Simulation details}

\subsection{Experimental data sets}

Total scattering structure factors that had been investigated by conventional RMC earlier \cite{Pusztai1999} were tested here: the X-ray structure factor of H$_2$O from Narten and Levy \cite{Narten1971} (this dataset will be denoted throughout this paper as XRD1) and the neutron structure factor of D$_2$O from Soper et al. \cite{Soper1997} (indicated as ND). Additionally, a recent X-ray diffraction result, the $S_{\mathrm{OO}}$ partial structure factor of Skinner and co-workers \cite{Skinner2013} (marked as XRD2) was considered. The ND structure function was modeled over the 1.1~\AA$^{-1}$ $\leq Q \leq $ 15~\AA$^{-1}$ regime. The weights of the partial functions in the ND total scattering function were 0.0919 (O-O), 0.4225 (O-D) and 0.4857 (D-D). XRD1 $S(Q)$ was considered for $Q$ values 1~\AA$^{-1}\leq Q \leq$ 16~\AA$^{-1}$, whereas the $S_{\mathrm{OO}}$ partial structure factor (XRD2) was fitted in the 0.975~\AA$^{-1}\leq Q \leq$ 25.85~\AA$^{-1}$ region.

In the first case (denoted as Case 1) the ND and the XRD1 data sets were fitted together, while in the second set of RMC simulations (marked as Case 2) the ND and XRD2 structure factors were approached simultaneously.

\subsection{Molecular dynamics simulations}

Molecular dynamics simulations have been performed by the GROMACS software package (version 5.1.1)\cite{GROMACS}. The initial particle configuration was obtained by placing 3333 water molecules in the cubic simulation box by the 'gmx solvate' program of the GROMACS package. The edge length of the simulation box was 46.5701~\AA{}, according to an atomic number density of 0.099~\AA$^{-3}$.

According to the SPC/E water potential \cite{Berendsen1987}, pairwise-additive non-bonded interactions have been used for the representation of dispersion and repulsion effects (in the form of the Lennard-Jones  (LJ) potential), as well as for the electrostatic interactions. In the SPC/E force field\cite{Berendsen1987}, which was chosen for this study, the charges are -0.8476$e$ and 0.4238$e$ for O and H atoms, respectively ($e$ is the elementary charge). The LJ $\sigma$ and $\epsilon$ parameters are 3.16557~\AA{} and 0.650194~kJ/mol for the oxygen atoms and 0 for hydrogens. Intramolecular distances are 1~\AA{} for O-H and 1.633~\AA{} for H-H pairs in the rigid molecules. 

Canonical NVT (constant number, volume and temperature) ensemble was applied at $T=$~295~K. The temperature was controlled by the Berendsen thermostat \cite{Berendsen1984}, with temperature coupling time $\tau _T =$~0.01~ps. The cutoffs for the Coulomb and van der Waals interactions were 10~\AA{}. 
The steepest-descent gradient method was used to reach the energy equilibrium.
The total energy has reached its minimum value in less than 100~ps. 
The total simulation time was 4000~ps, the time step was 2~fs; particle configurations for calculating averages have been collected in every 20~ps between 2000-4000~ps. MD results reported here were averaged over 100 time frames. The 'gmx~rdf' program was used to calculate the partial pair correlation functions from the collected MD configurations.

MD simulations were conducted with the flexible SPC/Fw water model \cite{Wu2006}, as well. In this model the LJ $\sigma$ and $\epsilon$ parameters are the same as in the SPC/E force field. The charges are -0.82$e$ and 0.41$e$ for O and H atoms, respectively. The flexibility is realized by harmonic bond stretching and angle bending potentials. The equilibrium O-H bond length is 1.012~\AA, the $k_b$ force constant is  443153~kJ mol$^{-1}$ nm$^{-2}$, the equilibrium H-O-H angle is 113.24~\textdegree, the $k_a$ force constant is 317.56~kJ mol$^{-1}$ rad$^{-2}$.
Details of the simulation were similar as before, but the time step was smaller (0.2~fs). Results from this simulation will be referred to as 'MD-flexible', while those from the preceding calculations (with SPC/E water model) will be denoted as 'MD-rigid'.

\subsection{Reverse Monte Carlo modeling}

RMC modeling is described in detail in Refs. \onlinecite{Evrard2005,Gereben2007,Gereben2012,McGreevy2001}. During a conventional RMC calculation particles are moved randomly in the simulation box and differences between experimental and model structural quantities are minimized. RMC may be used for any quantity that can be expressed from the atomic coordinates (e.g. structure factors from diffraction experiments, EXAFS signals, or model pair correlation functions). If the squared differences between experimental and calculated data sets decrease by the move of a particle then the move is accepted, otherwise it is only accepted with some probability.

In the present investigation, several different RMC calculations have been performed; they are summarized in Table \ref{tab:models}. 

\begin{table*}
 \caption{Reverse Monte Carlo calculations performed.}
 \label{tab:models}
 \begin{ruledtabular}
  \begin{tabular}{lllll}
   RMC calculation&Short name&Data sets&Starting configuration&Potential/g(r) data sets\\
   \hline
   RMC-FNC Case 1&FNC\_X1&ND+XRD1&MD-rigid&-\\
   RMC-FNC Case 2&FNC\_X2&ND+XRD2&MD-rigid&-\\
   RMC-FNC+g(r) Case 1&FNC\_g\_X1&ND+XRD1&MD-rigid& g(r) sets\\
   RMC-FNC+g(r) Case 2&FNC\_g\_X2&ND+XRD2&MD-rigid& g(r) sets\\
   RMC\_POT Case 1&POT\_X1&ND+XRD1&MD-rigid&SPC/Ef\\
   RMC\_POT Case 2&POT\_X2&ND+XRD2&MD-rigid&SPC/Ef\\
   RMC\_POT random&POT\_r&ND+XRD2&random&SPC/Ef\\
   RMC\_POT SPC/Fw&POT\_Fw&ND+XRD2&MD-flexible&SPC/Fw\\
  \end{tabular}
 \end{ruledtabular}
\end{table*}

Along with RMC calculations that use interatomic potential functions, traditional RMC modeling with Fixed Neighbour Constraints (FNC) has also been carried out for comparison. For these computations the RMC++ computer programme \cite{Gereben2007} has been applied (for an early application of such an approach for liquid water, see, e.g., Ref. \onlinecite{Pusztai1999}). In RMC++ molecules are kept together via FNC; no interatomic potentials are involved. The FNC method connects two hydrogen atoms and the central oxygen atom permanently via a simple neighbor list. Intramolecular distances are kept between minimum and maximum values, namely 0.95 to 1.03~\AA{} for (covalently bonded) O-H and 1.55 to 1.70~\AA{} for H-H (non-bonded intramolecular) pairs. Closest intermolecular approaches were 2.2~\AA{} for O-O, 1.5~\AA{} for O-H and 1.7~\AA{} for H-H pairs. The bin size was 0.05~\AA{}, and the maximum move of atoms was set to 0.05~\AA{}. The final MD-rigid configuration was taken as starting configuration for these simulations. Control parameters for experimental data sets are presented in Table \ref{tab:sigmas}.
These runs will be referred to as 'RMC-FNC' throughout this work.

\begin{table}
 \caption{\label{tab:sigmas} Final control parameters ($\sigma$ values) of input data sets and potential terms.}
 \begin{ruledtabular}
 
 \begin{tabular}{@{}ll}
  Data Set/Potential&$\sigma$\\ 
 \hline
 ND&0.003\\
 XRD1&0.003\\
 XRD2&0.005\\ 
 LJ&1.0\footnote{This value was 0.6 for POT\_Fw}\\
 Coulomb&1.0$^\text{a}$\\
 Bond&0.5\\
 Angle&0.5\\
 g$_\mathrm{O-O}$&0.04\\
 g$_\mathrm{O-H,intra}$&0.1\\
 g$_\mathrm{O-H,inter}$&0.04\\
 g$_\mathrm{H-H,intra}$&0.01\\
 g$_\mathrm{H-H,inter}$&0.04\\
 \end{tabular}
 \end{ruledtabular}
\end{table}

In an additional reference series of RMC calculations the FNC method has been used again and in addition, partial pair correlation functions ($g_{ij}(r)$) from MD simulations have been applied as '(quasi-)experimental data sets to fit'; these RMC runs will be denoted as 'RMC-FNC+g(r)'. A similar approach had been used and investigated previously in conjunction with several water models in Pusztai et al. \cite{Pusztai2008} and Steinczinger et al. \cite{Steinczinger2013}. $g_{ij}(r)$ curves were obtained here from molecular dynamics simulation with the SPC/E force field (MD-rigid). The control parameters for these data sets are also shown in Table \ref{tab:sigmas}; 
all other parameters were the same as before.

In the RMC\_POT method, exploitation of which is the genuinely novel element of this study, molecules are kept together via intramolecular ('bonded') potentials that are calculated similarly to that implemented in GROMACS \cite{GROMACS}. In general, the bond stretching interaction (between atoms $i$ and $j$) is taken into account as a harmonic potential:
\begin{equation}
 \label{eq:Vbond}
 V_b \left( r_{ij} \right) = \frac{1}{2} k_{ij}^b \left( r_{ij} - b_{ij} \right)^2
\end{equation} 
where $k_{ij}^b$ is the force constant, $b_{ij}$ is the equilibrium distance of the bonded pair, $r_{ij}$ is the actual distance of the  atoms in the bonded pair. The harmonic angle bending potential (between atoms $i$, $j$, and $k$, where atom $j$ is in the middle) is
\begin{equation}
 \label{eq:Vangle}
 V_a \left( \theta_{ijk} \right) = \frac{1}{2} k_{ijk}^a \left( \theta_{ijk} - \theta_{ijk}^0 \right)^2
\end{equation} 
where $k_{ij}^a$ is the force constant and $\theta_{ijk}^0$ is the equilibrium angle.

The non-bonded potential energy terms in these simulations are the Coulomb and the Lennard-Jones contributions. The Coulomb potential is
\begin{equation}
 \label{eq:Coulomb}
V_C \left( r_{ij} \right) = \frac{1}{4 \pi \epsilon_0} \frac{q_i q_j}{r_{ij}}
\end{equation} 
where $q_i$ and $q_j$ are the partial charges placed on atoms $i$ and $j$, $\epsilon_0$ is the vacuum permittivity.
The Lennard-Jones potential is
\begin{equation}
 \label{eq:LJ}
V_{LJ} \left( r_{ij} \right) = 4 \epsilon_{ij} \left( \left( \frac{\sigma_{ij}}{r_{ij}} \right)^{12} - \left( \frac{\sigma_{ij}}{r_{ij}} \right)^6 \right) 
\end{equation} 
where $\epsilon_{ij}$ and $\sigma_{ij}$ are the Lennard-Jones parameters applied for the $ij$ atom pair.

During RMC\_POT calculations differences between experimental and model $S(Q)$ functions are minimized together with the total potential energy. 
A relative weight ($\sigma$ parameter) is assigned to every potential related term. From the potential-related terms $\chi^2_{\mathrm{i}} = V_{\mathrm{i}} / \sigma_{\mathrm{i}}$ values are calculated and the sum of them is minimized ($i$ refers to the individual potential related components). After moving a particle, first the potential-related  terms are investigated and the move is accepted if the sum of the $\chi^2_{\mathrm{i}}$ terms has decreased. If it has increased then the move is accepted with some probability. Calculations related to conventional data sets (and geometrical constraints, if present) would be executed only if the move can be accepted based on potential energy.

Proper choice of the $\sigma$ parameters can warrant that information related to experimental data and interatomic potentials are taken into account in a balanced way. Relative weights of the bonding potential terms should be chosen so that molecules are kept together but they can be as flexible as diffraction data might require. Weights that are too strict do not allow the system to move around; on the other hand, too loose relative weights result in that the molecules break up. The ratio of the potential and experiment related weights should be chosen so that the fit to the experimental data sets is as good as possible, i.e., at least of similar quality as it may be achieved in an RMC++ calculation. 
As a result of RMC\_POT, final particle configurations will be compatible with the experimental data sets and the applied force field models simultaneously -- provided that these two things can be made compatible at all for a given system.

It should be noted here that as a consequence of the way molecules are handled in the RMC\_POT method, and of that movements are overwhelmingly atomic in RMC modeling in general, molecules in practice are always flexible (or perhaps better to say, never strictly rigid). Flexible water models are treated the same way as in an MD simulation. However, instead of the rigid SPC/E water model a modified variant of this model (referred to as SPC/Ef throughout this work) was effective during the present RMC\_POT simulations. It was realized in a similar way as in the work of Teleman and co-workers \cite{Teleman1987}: the basic idea is that a rigid model can be treated as a flexible model with infinitely strong force constants ($k_a=k_b=\infty$). In the SPC/Ef model every single parameter of the SPC/E model are maintained, but finite force constants are introduced. The force constant for bond stretching ($k_b$) was 463700~kJmol$^{-1}$nm$^{-2}$ and the angle bending force constant ($k_a$) was 383~kJmol$^{-1}$rad$^{-2}$, in alignment with Ref. \onlinecite{Teleman1987}. The exact values of the force constants are not critical, since the relative weights of the potential-related terms in the RMC\_POT scheme determine the significance of these 'intramolecular bonding' energy contributions to the total $\chi^2$ and thus, to the final particle configuration.

In the present study version 1.4 of RMC\_POT\cite{Gereben2012} has been made use of. 
Intermolecular O-O and O-H minimum distances were the same as for the RMC-FNC simulation runs (2.2 \AA{} and 1.5 \AA{}, respectively). For H-H pairs, inter- and intramolecular closest approaches cannot be defined separately, since intra- and intermolecular regions may overlap; considering this possibility, the H-H minimal interatomic distance was set at 1.4 \AA. Bin size, maximum moves of the particles and relative weights of the experimental data sets were the same as in the traditional RMC runs.

As a cross-check with a genuine flexible force field, RMC\_POT calculations have been performed applying the flexible SPC/Fw \cite{Wu2006}, along with the 'flexible' version of (the originally rigid) SPC/E, a.k.a. SPC/Ef (see above) potential. The final MD-flexible configuration was taken as starting configuration for the RMC\_POT runs performed with the SPC/Fw potential (this calculation is denoted as 'RMC\_POT SPC/Fw'). 
RMC\_POT simulations with SPC/Ef were started from the final MD-rigid configuration. The (possible) influence of the starting configuration was checked by a special simulation run: a random configuration of water molecules had been generated by placing water molecules randomly to the simulation box,
and this random initial configuration was used in this 'RMC\_POT random' run. (We note in passing that according to the best of our knowledge, no such test has been conducted for any other potential-related RMC-like method before.) This 'RMC\_POT random' calculation required more computational time (higher number of accepted moves) to reach an equilibrium, therefore it has only been performed for one combination of potential and experimental data sets. During the 'RMC\_POT random' simulation the SPC/Ef potential was applied and the ND and XRD2 experimental data sets were fitted.

Good values of the $\sigma$ parameters have resulted from several test runs, in which the number of accepted moves was about $10^6$. In the final runs the number of the accepted moves was around $2-8 \times 10^7$. The final $\sigma$ values are shown in Table \ref{tab:sigmas}.

\section{Results and discussion}

\begin{figure}
 \begin{center}
  \includegraphics[width=\columnwidth]{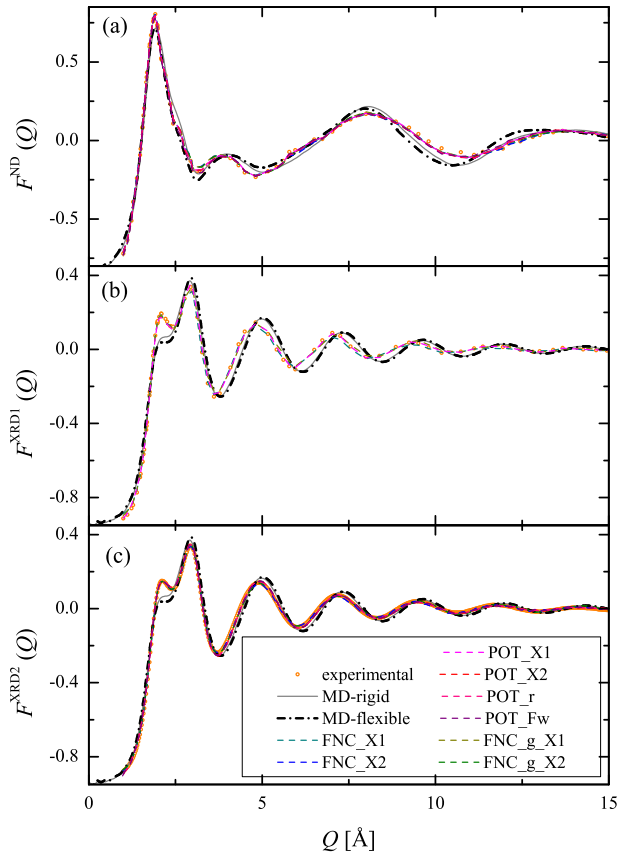} 
 \caption{Experimental, MD simulated and RMC total structure factors. (a) Neutron diffraction data of Soper \cite{Soper1997}; (b) X-ray diffraction data of Narten \cite{Narten1971}; (c) the recent O-O partial structure factor derived from X-ray data by Skinner \cite{Skinner2013}.}
 \label{fig:fits}
 \end{center}
\end{figure}

RMC and MD simulated total structure factors, along with their experimental counterparts, are shown in Figure \ref{fig:fits}. Agreement between any RMC calculation and experimental data is very good, whereas differences between MD simulations and experiment are well visible (although it is worth noting that both the rigid SPC/E and the flexible SPC/Fw potentials work remarkably well). Goodness-of-fit values ($R$-factors) are provided in Table \ref{tab:R-factors}: in accordance with visual inspection, $R$-factors also indicate similar fit qualities for all the RMC models, whereas differences between MD and experiment appear as 'magnified'.

\begin{table}
 \caption{\label{tab:R-factors} Goodness-of-fit values ($R$-factors) for each calculation.}
 \begin{ruledtabular}
 
 \begin{tabular}{@{}llll}
 &\multicolumn{3}{c}{$R$-factors (in \%)}\\
 &\\
 &ND&XRD1&XRD2\\
 \hline
 MD-rigid&15.4&19.7&13.5\\
 MD-flexible&13.6&24.4&18.5\\
 FNC\_X1&3.25&2.81&-\\
 FNC\_X2&3.27&-&3.15\\
 FNC\_g\_X1&4.20&4.14&-\\
 FNC\_g\_X2&4.14&-&4.20\\
 POT\_X1&3.52&3.54&-\\
 POT\_X2&3.34&-&4.34\\
 POT\_r&3.25&-&4.22\\
 POT\_Fw&3.70&-&4.47\\      
  \end{tabular}
 \end{ruledtabular}
\end{table}

\begin{figure}
 \includegraphics[width=\columnwidth]{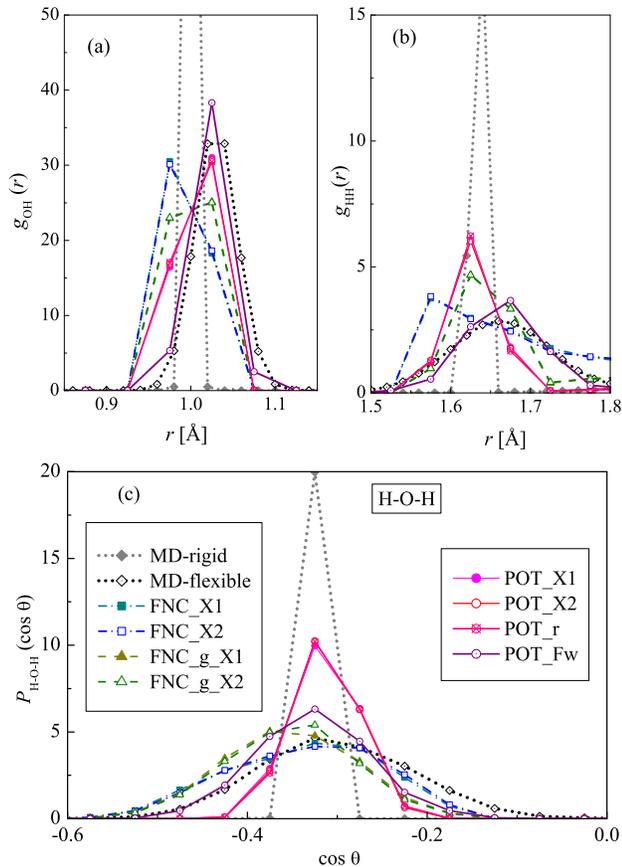}
 \caption{Intramolecular O-H bonding (a), H$\cdots$H non-bonding (b) distances and the distribution of the cosines of intramolecular H-O-H angles (c).}
 \label{fig:intra}
\end{figure}

Since in an earlier communication \cite{Pethes2015} some dispute appeared concerning intramolecular parameters that may be derived from diffraction experiments, here we show distributions of O-H (bonding) and H$\cdots$H (non-bonding) intramolecular distances, as well as of H-O-H bond angles, in Figure \ref{fig:intra}. Clearly, RMC\_POT with the SPC/Ef potential parameters leads to a molecular geometry that is rather similar to that obtained from molecular dynamics simulations (with the SPC/E potential), although there are slight differences: the average H$\cdots$H non-bonding distance is somewhat (by cca. 0.02 \AA{}) shorter as a result of RMC\_POT, and as a result, the average bond angle is very slightly smaller. The mean bond angle from all RMC\_POT calculations (regardless of the actual potential and starting configuration) appear to be identical, although the distribution of intramolecular H-O-H angles is significantly wider when the flexible SPC/Fw force field is applied. Calculations using the RMC-FNC approach provide significantly broader distributions for these intramolecular parameters. As each RMC calculation produced agreement with experimental data at the same (very high) level, it is established that the data used here allow for a diversity of the H$\cdots$H non-bonding intramolecular distance and of the bond angle as shown in Figure \ref{fig:intra}.

\begin{figure}
 \includegraphics[width=\columnwidth]{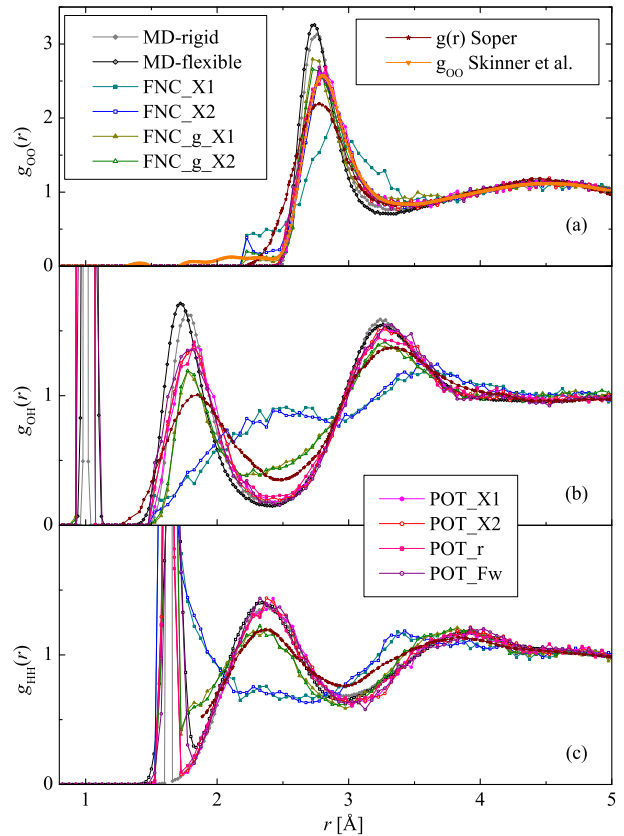}
 \caption{Partial pair correlation functions for each calculation, together with literature data from Soper \cite{Soper2015} and from Skinner et al. \cite{Skinner2013}.}
 \label{fig:ppcfs}
\end{figure}

\begin{figure}
 \includegraphics[width=\columnwidth]{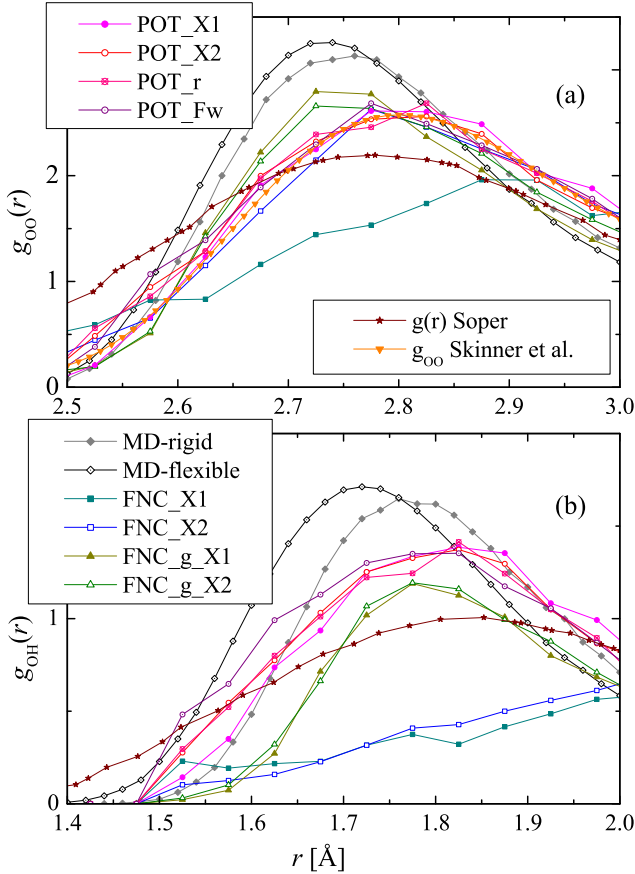}
 \caption{O-O and O-H partial pair correlation functions for each simulation, together with literature data from Soper \cite{Soper2015} and from Skinner et al. \cite{Skinner2013}; the focus is on the regions of the first intermolecular maxima.}
 \label{fig:ppcfs2}
\end{figure}

Partial pair correlation functions obtained for most of the (MD and RMC) simulations are presented in Figure 
\ref{fig:ppcfs}. The two XRD sets bring about very different O-O PPCF-s for the RMC-FNC calculations; also, none of the RMC-FNC simulations produced a separated O-H peak around the (assumed) hydrogen bonding distance (between cca. 1.75 and 1.95 \AA{}). Otherwise, the curves reflect, again, the diversity that one had to get used to in the literature of the structure of water (see, e.g., Refs. \onlinecite{Pusztai1999,Soper2015,Pethes2015,Soper1997}). O-O and O-H peak positions are gathered in table \ref{tab:peaks}: for the former, values between 2.75 and 2.9 \AA{}, while for the latter, those between 1.77 and 1.82 \AA{} have been found (if we disregard the most certainly unphysical O-H maxima at 2.6 \AA{} detected for the RMC-FNC runs). Again, the variously shaped functions in Figure \ref{fig:ppcfs} and \ref{fig:ppcfs2} and the different maximum position values in Table \ref{tab:peaks} that are related to RMC calculations are all equally consistent with the two sets of diffraction data: separation between them is only possible on the basis of external information. Note that the 'RMC\_POT random' calculation produced $g_{ij}(r)$-s that were indistinguishable from those that had resulted from the other RMC\_POT simulations. The intermolecular parts of each PPCF are identical for the 4 RMC\_POT calculations reported here (independently of the flexibility of the actual potential function used.).  

What is worth emphasizing is that RMC\_POT results from modeling the ND and XRD1 data sets have reproduced the 'experimental' $g_\mathrm{OO}(r)$ that belongs to the XRD2 data \cite{Skinner2013}. This is yet another sign of that data presented by Skinner et al. \cite{Skinner2013} are of high quality and that they may, indeed, be called as 'consensual' (although perhaps not quite as 'benchmark').

\begin{table}
 \caption{\label{tab:peaks} Positions of intramolecular O-H and H-H, and of the first intermolecular O-O and O-H maxima, as determined from the PPCF-s, for the different simulations (in \AA{}). Note that 'RMC-FNC' calculations have not provided any well distinguishable maximum for the hydrogen bonding distance (around 1.8 \AA{}).}
 \begin{ruledtabular}
 
 \begin{tabular}{@{}lllll}
 Simulation &O-H&H-H&O$\cdots$O&O$\cdots$H\\
 &intra&intra&inter&inter\\
 \hline
 MD-rigid&1.00&1.63&2.76&1.77\\
 MD-flexible&1.03&1.66&2.73&1.72\\
 FNC\_X1&1.00&1.58&2.9&2.6\\
 FNC\_X2&1.00&1.58&2.8&2.6\\
 FNC\_g\_X1&1.00&1.64&2.75&1.79\\
 FNC\_g\_X2&1.00&1.64&2.76&1.79\\
 POT\_X1&1.00&1.63&2.81&1.82\\
 POT\_X2&1.00&1.63&2.81&1.82\\
 POT\_r&1.00&1.63&2.80&1.82\\
 POT\_Fw&1.02&1.66&2.79&1.78\\
 Soper \cite{Soper2015}&&&2.77&1.84\\
 Skinner \cite{Skinner2013}&&&2.80&\\
  \end{tabular}
 \end{ruledtabular}
\end{table}

\begin{figure}
 \includegraphics[width=\columnwidth]{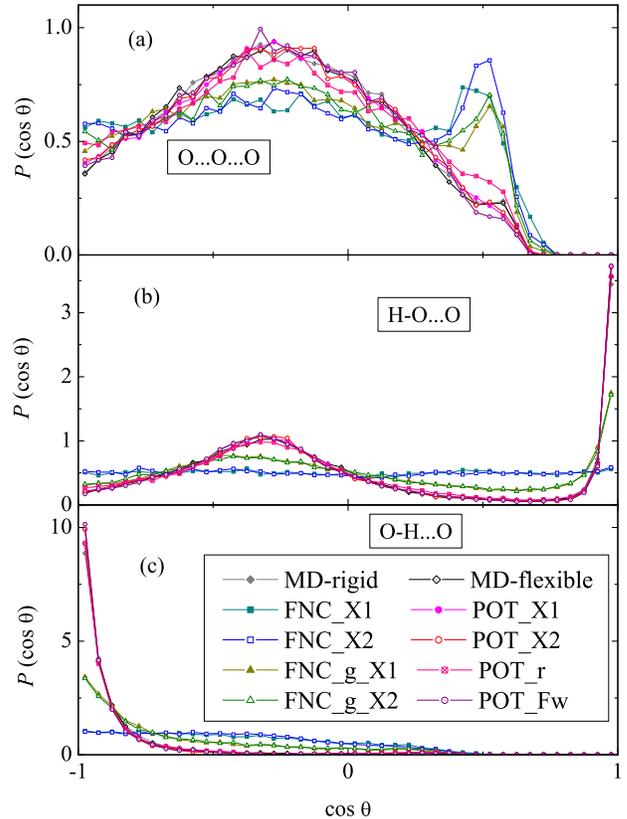}
 \caption{Distributions of the cosines of (a) O$\cdots$O$\cdots$O, (b) H-O$\cdots$O, and (c) O-H$\cdots$O angles.}
 \label{fig:angles}
\end{figure}

Distributions of intermolecular O$\cdots$O$\cdots$O, H-O$\cdots$O and (c) O-H$\cdots$O angles are shown in Figure \ref{fig:angles}: these are all characteristic to the local environment (including hydrogen bonding) of water molecules. Although the main features, the tetrahedral location of neighboring water molecules and the approximately straight hydrogen bonds, may be detected for each RMC calculations, the extents of these vary significantly. RMC\_POT results follow characteristics of the original molecular dynamics simulation nearly within the line thickness in the figures, whereas it is rather hard to detect the features in question on curves resulting from RMC-FNC calculations. (Again, 'RMC\_POT random' results were indistinguishable from those of the other RMC\_POT simulations; also, the originally rigid SPC/Ef and the flexible SPC/Fw potentials produce identical curves.)  

This behavior of RMC\_POT results is very encouraging: there seems to be no need for inventing extensive coordination constraints (c.f. Ref. \onlinecite{Pusztai1999}) in RMC any longer if we wish to obtain physically meaningful particle arrangements for liquid water. Interestingly, RMC\_POT provided slightly more regular tetrahedral local environment than even the original MD (cf. Figure \ref{fig:angles}, part (a)). 

A few words may be appropriate for a brief comparison with EPSR results on liquid water \cite{Soper2015,Soper2005}: there is a general agreement between RMC\_POT and EPSR in terms of the main characteristics, although some of the details may appear quite differently. Perhaps the most apparent of these differences concern the intramolecular structure, for which RMC\_POT seems to produce considerably narrower distributions for the O-H and H-H distances. The main reason behind may be the different handling of intramolecular contributions, cf. Ref. \cite{Soper2005}.

\section{Conclusions}
We have presented results from the first Reverse Monte Carlo study of liquid water in which a popular interatomic potential model, SPC/E \cite{Berendsen1987}, was applied explicitly. After some 'learning period', the calculations could be tuned so that the RMC\_POT algorithm was running with the same efficiency as usual for modeling studies without potentials; we therefore expect the extensive exploitation of this approach for water (with various input diffraction data sets and/or thermodynamic conditions) and aqueous solutions. 
Despite all of its attractive features, it has to be made clear that although RMC\_POT applies interatomic potentials explicitly, it still is a method of structural modeling and cannot be used to reliably calculate properties beyond structural ones.

The set of diffraction data applied here allowed the formation of local structural motifs in the RMC\_POT particle configurations that are hardly distinguishable from those produced by molecular dynamics simulations that use the same (SPC/E) water potential (cf. Figure \ref{fig:angles}). On the other hand, clear differences between MD and RMC\_POT findings were detected in terms of total structure factors and partial pair correlation functions. The very regular 'V-shape' of water molecules, the tetrahedral environment and straight hydrogen bond angles reflect the present 'collective wisdom' concerning the structure of liquid water: it is therefore rather comforting that the present study is able to offer large 3D particle configurations that, in addition, are fully consistent with diffraction data. These statements have proven to be valid independently of the initial particle configuration, as well as of the actual variant (rigid or flexible) of the SPC/E force field.

Even though reference (FNC-based) RMC calculations reproduced experimental diffraction data at the same (very high) level as RMC\_POT, and much better than molecular dynamics simulations, the outcome from these runs is less attractive: the molecular shapes are less uniform and hydrogen bond angles take values that are sometimes far from the ideal 180 degrees. Still, these arrangements are fully consistent with diffraction data considered here -- they just tend to be the most disordered ones of those that may be called 'realistic' (cf. Ref. \onlinecite{Pethes2015}).

\begin{acknowledgments}
The authors acknowledge financial support from the National Research, Development and Innovation Office of Hungary (NKFIH), under grant No. SNN 116198.
\end{acknowledgments}

\bibliography{h2o-rmcpot}

\begin{thebibliography}{43}%
\makeatletter
\providecommand \@ifxundefined [1]{%
 \@ifx{#1\undefined}
}%
\providecommand \@ifnum [1]{%
 \ifnum #1\expandafter \@firstoftwo
 \else \expandafter \@secondoftwo
 \fi
}%
\providecommand \@ifx [1]{%
 \ifx #1\expandafter \@firstoftwo
 \else \expandafter \@secondoftwo
 \fi
}%
\providecommand \natexlab [1]{#1}%
\providecommand \enquote  [1]{``#1''}%
\providecommand \bibnamefont  [1]{#1}%
\providecommand \bibfnamefont [1]{#1}%
\providecommand \citenamefont [1]{#1}%
\providecommand \href@noop [0]{\@secondoftwo}%
\providecommand \href [0]{\begingroup \@sanitize@url \@href}%
\providecommand \@href[1]{\@@startlink{#1}\@@href}%
\providecommand \@@href[1]{\endgroup#1\@@endlink}%
\providecommand \@sanitize@url [0]{\catcode `\\12\catcode `\$12\catcode
  `\&12\catcode `\#12\catcode `\^12\catcode `\_12\catcode `\%12\relax}%
\providecommand \@@startlink[1]{}%
\providecommand \@@endlink[0]{}%
\providecommand \url  [0]{\begingroup\@sanitize@url \@url }%
\providecommand \@url [1]{\endgroup\@href {#1}{\urlprefix }}%
\providecommand \urlprefix  [0]{URL }%
\providecommand \Eprint [0]{\href }%
\providecommand \doibase [0]{http://dx.doi.org/}%
\providecommand \selectlanguage [0]{\@gobble}%
\providecommand \bibinfo  [0]{\@secondoftwo}%
\providecommand \bibfield  [0]{\@secondoftwo}%
\providecommand \translation [1]{[#1]}%
\providecommand \BibitemOpen [0]{}%
\providecommand \bibitemStop [0]{}%
\providecommand \bibitemNoStop [0]{.\EOS\space}%
\providecommand \EOS [0]{\spacefactor3000\relax}%
\providecommand \BibitemShut  [1]{\csname bibitem#1\endcsname}%
\let\auto@bib@innerbib\@empty
\bibitem [{Note1()}]{Note1}%
  \BibitemOpen
  \bibinfo {note} {See e. g. http://www1.lsbu.ac.uk/water/.}\BibitemShut
  {Stop}%
\bibitem [{\citenamefont {Wernet}\ \emph {et~al.}(2004)\citenamefont {Wernet},
  \citenamefont {Nordlund}, \citenamefont {Bergmann}, \citenamefont
  {Cavalleri}, \citenamefont {Odelius}, \citenamefont {Ogasawara},
  \citenamefont {N\"{a}slund}, \citenamefont {Hirsch}, \citenamefont
  {Ojam\"{a}e}, \citenamefont {Glatzel}, \citenamefont {Pettersson},\ and\
  \citenamefont {Nilsson}}]{WernetScience2004}%
  \BibitemOpen
  \bibfield  {author} {\bibinfo {author} {\bibfnamefont {P.}~\bibnamefont
  {Wernet}}, \bibinfo {author} {\bibfnamefont {D.}~\bibnamefont {Nordlund}},
  \bibinfo {author} {\bibfnamefont {U.}~\bibnamefont {Bergmann}}, \bibinfo
  {author} {\bibfnamefont {M.}~\bibnamefont {Cavalleri}}, \bibinfo {author}
  {\bibfnamefont {M.}~\bibnamefont {Odelius}}, \bibinfo {author} {\bibfnamefont
  {H.}~\bibnamefont {Ogasawara}}, \bibinfo {author} {\bibfnamefont {L.~{\AA}.}\
  \bibnamefont {N\"{a}slund}}, \bibinfo {author} {\bibfnamefont {T.~K.}\
  \bibnamefont {Hirsch}}, \bibinfo {author} {\bibfnamefont {L.}~\bibnamefont
  {Ojam\"{a}e}}, \bibinfo {author} {\bibfnamefont {P.}~\bibnamefont {Glatzel}},
  \bibinfo {author} {\bibfnamefont {L.~G.~M.}\ \bibnamefont {Pettersson}}, \
  and\ \bibinfo {author} {\bibfnamefont {A.}~\bibnamefont {Nilsson}},\ }\href
  {\doibase 10.1126/science.1096205} {\bibfield  {journal} {\bibinfo  {journal}
  {Science}\ }\textbf {\bibinfo {volume} {304}},\ \bibinfo {pages} {995}
  (\bibinfo {year} {2004})}\BibitemShut {NoStop}%
\bibitem [{\citenamefont {Head-Gordon}\ and\ \citenamefont
  {Johnson}(2006)}]{Head-Gordon2006}%
  \BibitemOpen
  \bibfield  {author} {\bibinfo {author} {\bibfnamefont {T.}~\bibnamefont
  {Head-Gordon}}\ and\ \bibinfo {author} {\bibfnamefont {M.~E.}\ \bibnamefont
  {Johnson}},\ }\href {\doibase 10.1073/pnas.0510593103} {\bibfield  {journal}
  {\bibinfo  {journal} {Proc. Natl. Acad. Sci. U.S.A.}\ }\textbf {\bibinfo
  {volume} {103}},\ \bibinfo {pages} {7973} (\bibinfo {year}
  {2006})}\BibitemShut {NoStop}%
\bibitem [{\citenamefont {Leetmaa}\ \emph {et~al.}(2008)\citenamefont
  {Leetmaa}, \citenamefont {Wikfeldt}, \citenamefont {Ljungberg}, \citenamefont
  {Odelius}, \citenamefont {Swenson}, \citenamefont {Nilsson},\ and\
  \citenamefont {Pettersson}}]{Leetmaa2008}%
  \BibitemOpen
  \bibfield  {author} {\bibinfo {author} {\bibfnamefont {M.}~\bibnamefont
  {Leetmaa}}, \bibinfo {author} {\bibfnamefont {K.~T.}\ \bibnamefont
  {Wikfeldt}}, \bibinfo {author} {\bibfnamefont {M.~P.}\ \bibnamefont
  {Ljungberg}}, \bibinfo {author} {\bibfnamefont {M.}~\bibnamefont {Odelius}},
  \bibinfo {author} {\bibfnamefont {J.}~\bibnamefont {Swenson}}, \bibinfo
  {author} {\bibfnamefont {A.}~\bibnamefont {Nilsson}}, \ and\ \bibinfo
  {author} {\bibfnamefont {L.~G.~M.}\ \bibnamefont {Pettersson}},\ }\href
  {\doibase 10.1063/1.2968550} {\bibfield  {journal} {\bibinfo  {journal} {J.
  Chem. Phys.}\ }\textbf {\bibinfo {volume} {129}},\ \bibinfo {pages} {084502}
  (\bibinfo {year} {2008})}\BibitemShut {NoStop}%
\bibitem [{\citenamefont {Pusztai}(1999)}]{Pusztai1999}%
  \BibitemOpen
  \bibfield  {author} {\bibinfo {author} {\bibfnamefont {L.}~\bibnamefont
  {Pusztai}},\ }\href {\doibase 10.1103/PhysRevB.60.11851} {\bibfield
  {journal} {\bibinfo  {journal} {Phys. Rev. B}\ }\textbf {\bibinfo {volume}
  {60}},\ \bibinfo {pages} {11851} (\bibinfo {year} {1999})}\BibitemShut
  {NoStop}%
\bibitem [{\citenamefont {Skinner}\ \emph {et~al.}(2013)\citenamefont
  {Skinner}, \citenamefont {Huang}, \citenamefont {Schlesinger}, \citenamefont
  {Pettersson}, \citenamefont {Nilsson},\ and\ \citenamefont
  {Benmore}}]{Skinner2013}%
  \BibitemOpen
  \bibfield  {author} {\bibinfo {author} {\bibfnamefont {L.~B.}\ \bibnamefont
  {Skinner}}, \bibinfo {author} {\bibfnamefont {C.}~\bibnamefont {Huang}},
  \bibinfo {author} {\bibfnamefont {D.}~\bibnamefont {Schlesinger}}, \bibinfo
  {author} {\bibfnamefont {L.~G.~M.}\ \bibnamefont {Pettersson}}, \bibinfo
  {author} {\bibfnamefont {A.}~\bibnamefont {Nilsson}}, \ and\ \bibinfo
  {author} {\bibfnamefont {C.~J.}\ \bibnamefont {Benmore}},\ }\href {\doibase
  10.1063/1.4790861} {\bibfield  {journal} {\bibinfo  {journal} {J. Chem.
  Phys.}\ }\textbf {\bibinfo {volume} {138}},\ \bibinfo {pages} {074506}
  (\bibinfo {year} {2013})}\BibitemShut {NoStop}%
\bibitem [{\citenamefont {Hura}\ \emph {et~al.}(2003)\citenamefont {Hura},
  \citenamefont {Russo}, \citenamefont {Glaeser}, \citenamefont {Head-Gordon},
  \citenamefont {Krack},\ and\ \citenamefont {Parrinello}}]{Hura2003}%
  \BibitemOpen
  \bibfield  {author} {\bibinfo {author} {\bibfnamefont {G.}~\bibnamefont
  {Hura}}, \bibinfo {author} {\bibfnamefont {D.}~\bibnamefont {Russo}},
  \bibinfo {author} {\bibfnamefont {R.~M.}\ \bibnamefont {Glaeser}}, \bibinfo
  {author} {\bibfnamefont {T.}~\bibnamefont {Head-Gordon}}, \bibinfo {author}
  {\bibfnamefont {M.}~\bibnamefont {Krack}}, \ and\ \bibinfo {author}
  {\bibfnamefont {M.}~\bibnamefont {Parrinello}},\ }\href {\doibase
  10.1039/b301481a} {\bibfield  {journal} {\bibinfo  {journal} {Phys. Chem.
  Chem. Phys.}\ }\textbf {\bibinfo {volume} {5}},\ \bibinfo {pages} {1981}
  (\bibinfo {year} {2003})}\BibitemShut {NoStop}%
\bibitem [{\citenamefont {Narten}\ and\ \citenamefont
  {Levy}(1971)}]{Narten1971}%
  \BibitemOpen
  \bibfield  {author} {\bibinfo {author} {\bibfnamefont {A.~H.}\ \bibnamefont
  {Narten}}\ and\ \bibinfo {author} {\bibfnamefont {H.~A.}\ \bibnamefont
  {Levy}},\ }\href {\doibase 10.1063/1.1676403} {\bibfield  {journal} {\bibinfo
   {journal} {J. Chem. Phys.}\ }\textbf {\bibinfo {volume} {55}},\ \bibinfo
  {pages} {2263} (\bibinfo {year} {1971})}\BibitemShut {NoStop}%
\bibitem [{\citenamefont {Huang}\ \emph {et~al.}(2009)\citenamefont {Huang},
  \citenamefont {Wikfeldt}, \citenamefont {Tokushima}, \citenamefont
  {Nordlund}, \citenamefont {Harada}, \citenamefont {Bergmann}, \citenamefont
  {Niebuhr}, \citenamefont {Weiss}, \citenamefont {Horikawa}, \citenamefont
  {Leetmaa}, \citenamefont {Ljungberg}, \citenamefont {Takahashi},
  \citenamefont {Lenz}, \citenamefont {Ojamae}, \citenamefont {Lyubartsev},
  \citenamefont {Shin}, \citenamefont {Pettersson},\ and\ \citenamefont
  {Nilsson}}]{Huang2009}%
  \BibitemOpen
  \bibfield  {author} {\bibinfo {author} {\bibfnamefont {C.}~\bibnamefont
  {Huang}}, \bibinfo {author} {\bibfnamefont {K.~T.}\ \bibnamefont {Wikfeldt}},
  \bibinfo {author} {\bibfnamefont {T.}~\bibnamefont {Tokushima}}, \bibinfo
  {author} {\bibfnamefont {D.}~\bibnamefont {Nordlund}}, \bibinfo {author}
  {\bibfnamefont {Y.}~\bibnamefont {Harada}}, \bibinfo {author} {\bibfnamefont
  {U.}~\bibnamefont {Bergmann}}, \bibinfo {author} {\bibfnamefont
  {M.}~\bibnamefont {Niebuhr}}, \bibinfo {author} {\bibfnamefont {T.~M.}\
  \bibnamefont {Weiss}}, \bibinfo {author} {\bibfnamefont {Y.}~\bibnamefont
  {Horikawa}}, \bibinfo {author} {\bibfnamefont {M.}~\bibnamefont {Leetmaa}},
  \bibinfo {author} {\bibfnamefont {M.~P.}\ \bibnamefont {Ljungberg}}, \bibinfo
  {author} {\bibfnamefont {O.}~\bibnamefont {Takahashi}}, \bibinfo {author}
  {\bibfnamefont {A.}~\bibnamefont {Lenz}}, \bibinfo {author} {\bibfnamefont
  {L.}~\bibnamefont {Ojamae}}, \bibinfo {author} {\bibfnamefont {A.~P.}\
  \bibnamefont {Lyubartsev}}, \bibinfo {author} {\bibfnamefont
  {S.}~\bibnamefont {Shin}}, \bibinfo {author} {\bibfnamefont {L.~G.~M.}\
  \bibnamefont {Pettersson}}, \ and\ \bibinfo {author} {\bibfnamefont
  {A.}~\bibnamefont {Nilsson}},\ }\href {\doibase 10.1073/pnas.0904743106}
  {\bibfield  {journal} {\bibinfo  {journal} {Proc. Natl. Acad. Sci. U.S.A.}\
  }\textbf {\bibinfo {volume} {106}},\ \bibinfo {pages} {15214} (\bibinfo
  {year} {2009})}\BibitemShut {NoStop}%
\bibitem [{\citenamefont {Clark}\ \emph {et~al.}(2010)\citenamefont {Clark},
  \citenamefont {Hura}, \citenamefont {Teixeira}, \citenamefont {Soper},\ and\
  \citenamefont {Head-Gordon}}]{Clark2010PNAS}%
  \BibitemOpen
  \bibfield  {author} {\bibinfo {author} {\bibfnamefont {G.~N.~I.}\
  \bibnamefont {Clark}}, \bibinfo {author} {\bibfnamefont {G.~L.}\ \bibnamefont
  {Hura}}, \bibinfo {author} {\bibfnamefont {J.}~\bibnamefont {Teixeira}},
  \bibinfo {author} {\bibfnamefont {A.~K.}\ \bibnamefont {Soper}}, \ and\
  \bibinfo {author} {\bibfnamefont {T.}~\bibnamefont {Head-Gordon}},\ }\href
  {\doibase 10.1073/pnas.1006599107} {\bibfield  {journal} {\bibinfo  {journal}
  {Proc. Natl. Acad. Sci. U.S.A.}\ }\textbf {\bibinfo {volume} {107}},\
  \bibinfo {pages} {14003} (\bibinfo {year} {2010})}\BibitemShut {NoStop}%
\bibitem [{\citenamefont {Soper}(2015)}]{Soper2015}%
  \BibitemOpen
  \bibfield  {author} {\bibinfo {author} {\bibfnamefont {A.~K.}\ \bibnamefont
  {Soper}},\ }\href {\doibase 10.1021/jp509909w} {\bibfield  {journal}
  {\bibinfo  {journal} {J. Phys. Chem. B}\ }\textbf {\bibinfo {volume} {119}},\
  \bibinfo {pages} {9244} (\bibinfo {year} {2015})}\BibitemShut {NoStop}%
\bibitem [{\citenamefont {Zeidler}\ \emph {et~al.}(2012)\citenamefont
  {Zeidler}, \citenamefont {Salmon}, \citenamefont {Fischer}, \citenamefont
  {Neuefeind}, \citenamefont {{Mike Simonson}},\ and\ \citenamefont
  {Markland}}]{Zeidler-Salmon2012}%
  \BibitemOpen
  \bibfield  {author} {\bibinfo {author} {\bibfnamefont {A.}~\bibnamefont
  {Zeidler}}, \bibinfo {author} {\bibfnamefont {P.~S.}\ \bibnamefont {Salmon}},
  \bibinfo {author} {\bibfnamefont {H.~E.}\ \bibnamefont {Fischer}}, \bibinfo
  {author} {\bibfnamefont {J.~C.}\ \bibnamefont {Neuefeind}}, \bibinfo {author}
  {\bibfnamefont {J.}~\bibnamefont {{Mike Simonson}}}, \ and\ \bibinfo {author}
  {\bibfnamefont {T.~E.}\ \bibnamefont {Markland}},\ }\href {\doibase
  10.1088/0953-8984/24/28/284126} {\bibfield  {journal} {\bibinfo  {journal}
  {J. Phys.: Condens. Matter}\ }\textbf {\bibinfo {volume} {24}},\ \bibinfo
  {pages} {284126} (\bibinfo {year} {2012})}\BibitemShut {NoStop}%
\bibitem [{\citenamefont {Narten}, \citenamefont {Thiessen},\ and\
  \citenamefont {Blum}(1982)}]{Narten1982}%
  \BibitemOpen
  \bibfield  {author} {\bibinfo {author} {\bibfnamefont {A.~H.}\ \bibnamefont
  {Narten}}, \bibinfo {author} {\bibfnamefont {W.~E.}\ \bibnamefont
  {Thiessen}}, \ and\ \bibinfo {author} {\bibfnamefont {L.}~\bibnamefont
  {Blum}},\ }\href {\doibase 10.1126/science.217.4564.1033} {\bibfield
  {journal} {\bibinfo  {journal} {Science}\ }\textbf {\bibinfo {volume}
  {217}},\ \bibinfo {pages} {1033} (\bibinfo {year} {1982})}\BibitemShut
  {NoStop}%
\bibitem [{\citenamefont {Myneni}\ \emph {et~al.}(2002)\citenamefont {Myneni},
  \citenamefont {Luo}, \citenamefont {N\"{a}slund}, \citenamefont {Cavalleri},
  \citenamefont {Ojam\"{a}e}, \citenamefont {Ogasawara}, \citenamefont
  {Pelmenschikov}, \citenamefont {Wernet}, \citenamefont {V\"{a}terlein},
  \citenamefont {Heske}, \citenamefont {Hussain}, \citenamefont {Pettersson},\
  and\ \citenamefont {Nilsson}}]{Myneni2002}%
  \BibitemOpen
  \bibfield  {author} {\bibinfo {author} {\bibfnamefont {S.}~\bibnamefont
  {Myneni}}, \bibinfo {author} {\bibfnamefont {Y.}~\bibnamefont {Luo}},
  \bibinfo {author} {\bibfnamefont {L.~{\AA}.}\ \bibnamefont {N\"{a}slund}},
  \bibinfo {author} {\bibfnamefont {M.}~\bibnamefont {Cavalleri}}, \bibinfo
  {author} {\bibfnamefont {L.}~\bibnamefont {Ojam\"{a}e}}, \bibinfo {author}
  {\bibfnamefont {H.}~\bibnamefont {Ogasawara}}, \bibinfo {author}
  {\bibfnamefont {A.}~\bibnamefont {Pelmenschikov}}, \bibinfo {author}
  {\bibfnamefont {P.}~\bibnamefont {Wernet}}, \bibinfo {author} {\bibfnamefont
  {P.}~\bibnamefont {V\"{a}terlein}}, \bibinfo {author} {\bibfnamefont
  {C.}~\bibnamefont {Heske}}, \bibinfo {author} {\bibfnamefont
  {Z.}~\bibnamefont {Hussain}}, \bibinfo {author} {\bibfnamefont {L.~G.~M.}\
  \bibnamefont {Pettersson}}, \ and\ \bibinfo {author} {\bibfnamefont
  {A.}~\bibnamefont {Nilsson}},\ }\href {\doibase 10.1088/0953-8984/14/8/106}
  {\bibfield  {journal} {\bibinfo  {journal} {J. Phys.: Condens. Matter}\
  }\textbf {\bibinfo {volume} {14}},\ \bibinfo {pages} {L213} (\bibinfo {year}
  {2002})}\BibitemShut {NoStop}%
\bibitem [{\citenamefont {Smith}\ \emph {et~al.}(2004)\citenamefont {Smith},
  \citenamefont {Cappa}, \citenamefont {Wilson}, \citenamefont {Messer},
  \citenamefont {Cohen},\ and\ \citenamefont {Saykally}}]{Smith2004}%
  \BibitemOpen
  \bibfield  {author} {\bibinfo {author} {\bibfnamefont {J.~D.}\ \bibnamefont
  {Smith}}, \bibinfo {author} {\bibfnamefont {C.~D.}\ \bibnamefont {Cappa}},
  \bibinfo {author} {\bibfnamefont {K.~R.}\ \bibnamefont {Wilson}}, \bibinfo
  {author} {\bibfnamefont {B.~M.}\ \bibnamefont {Messer}}, \bibinfo {author}
  {\bibfnamefont {R.~C.}\ \bibnamefont {Cohen}}, \ and\ \bibinfo {author}
  {\bibfnamefont {R.~J.}\ \bibnamefont {Saykally}},\ }\href {\doibase
  10.1126/science.1102560} {\bibfield  {journal} {\bibinfo  {journal}
  {Science}\ }\textbf {\bibinfo {volume} {306}},\ \bibinfo {pages} {851}
  (\bibinfo {year} {2004})}\BibitemShut {NoStop}%
\bibitem [{\citenamefont {Tokushima}\ \emph {et~al.}(2008)\citenamefont
  {Tokushima}, \citenamefont {Harada}, \citenamefont {Takahashi}, \citenamefont
  {Senba}, \citenamefont {Ohashi}, \citenamefont {Pettersson}, \citenamefont
  {Nilsson},\ and\ \citenamefont {Shin}}]{Tokushima2008}%
  \BibitemOpen
  \bibfield  {author} {\bibinfo {author} {\bibfnamefont {T.}~\bibnamefont
  {Tokushima}}, \bibinfo {author} {\bibfnamefont {Y.}~\bibnamefont {Harada}},
  \bibinfo {author} {\bibfnamefont {O.}~\bibnamefont {Takahashi}}, \bibinfo
  {author} {\bibfnamefont {Y.}~\bibnamefont {Senba}}, \bibinfo {author}
  {\bibfnamefont {H.}~\bibnamefont {Ohashi}}, \bibinfo {author} {\bibfnamefont
  {L.~G.~M.}\ \bibnamefont {Pettersson}}, \bibinfo {author} {\bibfnamefont
  {A.}~\bibnamefont {Nilsson}}, \ and\ \bibinfo {author} {\bibfnamefont
  {S.}~\bibnamefont {Shin}},\ }\href {\doibase 10.1016/j.cplett.2008.04.077}
  {\bibfield  {journal} {\bibinfo  {journal} {Chem. Phys. Lett.}\ }\textbf
  {\bibinfo {volume} {460}},\ \bibinfo {pages} {387} (\bibinfo {year}
  {2008})}\BibitemShut {NoStop}%
\bibitem [{\citenamefont {Fuchs}\ \emph {et~al.}(2008)\citenamefont {Fuchs},
  \citenamefont {Zharnikov}, \citenamefont {Weinhardt}, \citenamefont {Blum},
  \citenamefont {Weigand}, \citenamefont {Zubavichus}, \citenamefont {B\"{a}r},
  \citenamefont {Maier}, \citenamefont {Denlinger}, \citenamefont {Heske},
  \citenamefont {Grunze},\ and\ \citenamefont {Umbach}}]{Fuchs2008}%
  \BibitemOpen
  \bibfield  {author} {\bibinfo {author} {\bibfnamefont {O.}~\bibnamefont
  {Fuchs}}, \bibinfo {author} {\bibfnamefont {M.}~\bibnamefont {Zharnikov}},
  \bibinfo {author} {\bibfnamefont {L.}~\bibnamefont {Weinhardt}}, \bibinfo
  {author} {\bibfnamefont {M.}~\bibnamefont {Blum}}, \bibinfo {author}
  {\bibfnamefont {M.}~\bibnamefont {Weigand}}, \bibinfo {author} {\bibfnamefont
  {Y.}~\bibnamefont {Zubavichus}}, \bibinfo {author} {\bibfnamefont
  {M.}~\bibnamefont {B\"{a}r}}, \bibinfo {author} {\bibfnamefont
  {F.}~\bibnamefont {Maier}}, \bibinfo {author} {\bibfnamefont {J.~D.}\
  \bibnamefont {Denlinger}}, \bibinfo {author} {\bibfnamefont {C.}~\bibnamefont
  {Heske}}, \bibinfo {author} {\bibfnamefont {M.}~\bibnamefont {Grunze}}, \
  and\ \bibinfo {author} {\bibfnamefont {E.}~\bibnamefont {Umbach}},\ }\href
  {\doibase 10.1103/PhysRevLett.100.027801} {\bibfield  {journal} {\bibinfo
  {journal} {Phys. Rev. Lett.}\ }\textbf {\bibinfo {volume} {100}},\ \bibinfo
  {pages} {027801} (\bibinfo {year} {2008})}\BibitemShut {NoStop}%
\bibitem [{\citenamefont {Zeidler}\ \emph {et~al.}(2011)\citenamefont
  {Zeidler}, \citenamefont {Salmon}, \citenamefont {Fischer}, \citenamefont
  {Neuefeind}, \citenamefont {Simonson}, \citenamefont {Lemmel}, \citenamefont
  {Rauch},\ and\ \citenamefont {Markland}}]{Zeidler-Salmon2011}%
  \BibitemOpen
  \bibfield  {author} {\bibinfo {author} {\bibfnamefont {A.}~\bibnamefont
  {Zeidler}}, \bibinfo {author} {\bibfnamefont {P.~S.}\ \bibnamefont {Salmon}},
  \bibinfo {author} {\bibfnamefont {H.~E.}\ \bibnamefont {Fischer}}, \bibinfo
  {author} {\bibfnamefont {J.~C.}\ \bibnamefont {Neuefeind}}, \bibinfo {author}
  {\bibfnamefont {J.~M.}\ \bibnamefont {Simonson}}, \bibinfo {author}
  {\bibfnamefont {H.}~\bibnamefont {Lemmel}}, \bibinfo {author} {\bibfnamefont
  {H.}~\bibnamefont {Rauch}}, \ and\ \bibinfo {author} {\bibfnamefont {T.~E.}\
  \bibnamefont {Markland}},\ }\href {\doibase 10.1103/PhysRevLett.107.145501}
  {\bibfield  {journal} {\bibinfo  {journal} {Phys. Rev. Lett.}\ }\textbf
  {\bibinfo {volume} {107}},\ \bibinfo {pages} {145501} (\bibinfo {year}
  {2011})}\BibitemShut {NoStop}%
\bibitem [{\citenamefont {Temleitner}\ \emph {et~al.}(2015)\citenamefont
  {Temleitner}, \citenamefont {Stunault}, \citenamefont {Cuello},\ and\
  \citenamefont {Pusztai}}]{Temleitner2015}%
  \BibitemOpen
  \bibfield  {author} {\bibinfo {author} {\bibfnamefont {L.}~\bibnamefont
  {Temleitner}}, \bibinfo {author} {\bibfnamefont {A.}~\bibnamefont
  {Stunault}}, \bibinfo {author} {\bibfnamefont {G.~J.}\ \bibnamefont
  {Cuello}}, \ and\ \bibinfo {author} {\bibfnamefont {L.}~\bibnamefont
  {Pusztai}},\ }\href {\doibase 10.1103/PhysRevB.92.014201} {\bibfield
  {journal} {\bibinfo  {journal} {Phys. Rev. B}\ }\textbf {\bibinfo {volume}
  {92}},\ \bibinfo {pages} {014201} (\bibinfo {year} {2015})}\BibitemShut
  {NoStop}%
\bibitem [{\citenamefont {Barker}\ and\ \citenamefont
  {Watts}(1969)}]{Barker1969}%
  \BibitemOpen
  \bibfield  {author} {\bibinfo {author} {\bibfnamefont {J.~A.}\ \bibnamefont
  {Barker}}\ and\ \bibinfo {author} {\bibfnamefont {R.~O.}\ \bibnamefont
  {Watts}},\ }\href {\doibase 10.1016/0009-2614(69)80119-3} {\bibfield
  {journal} {\bibinfo  {journal} {Chem. Phys. Lett.}\ }\textbf {\bibinfo
  {volume} {3}},\ \bibinfo {pages} {144} (\bibinfo {year} {1969})}\BibitemShut
  {NoStop}%
\bibitem [{\citenamefont {Guillot}(2002)}]{Guillot2002}%
  \BibitemOpen
  \bibfield  {author} {\bibinfo {author} {\bibfnamefont {B.}~\bibnamefont
  {Guillot}},\ }\href {\doibase 10.1016/S0167-7322(02)00094-6} {\bibfield
  {journal} {\bibinfo  {journal} {J. Mol. Liq.}\ }\textbf {\bibinfo {volume}
  {101}},\ \bibinfo {pages} {219} (\bibinfo {year} {2002})}\BibitemShut
  {NoStop}%
\bibitem [{\citenamefont {Szalewicz}, \citenamefont {Leforestier},\ and\
  \citenamefont {van~der Avoird}(2009)}]{Szalewicz2009}%
  \BibitemOpen
  \bibfield  {author} {\bibinfo {author} {\bibfnamefont {K.}~\bibnamefont
  {Szalewicz}}, \bibinfo {author} {\bibfnamefont {C.}~\bibnamefont
  {Leforestier}}, \ and\ \bibinfo {author} {\bibfnamefont {A.}~\bibnamefont
  {van~der Avoird}},\ }\href {\doibase 10.1016/j.cplett.2009.09.029} {\bibfield
   {journal} {\bibinfo  {journal} {Chem. Phys. Lett.}\ }\textbf {\bibinfo
  {volume} {482}},\ \bibinfo {pages} {1} (\bibinfo {year} {2009})}\BibitemShut
  {NoStop}%
\bibitem [{\citenamefont {Jorgensen}\ and\ \citenamefont
  {Tirado-Rives}(2005)}]{Jorgensen2005}%
  \BibitemOpen
  \bibfield  {author} {\bibinfo {author} {\bibfnamefont {W.~L.}\ \bibnamefont
  {Jorgensen}}\ and\ \bibinfo {author} {\bibfnamefont {J.}~\bibnamefont
  {Tirado-Rives}},\ }\href {\doibase 10.1073/pnas.0408037102} {\bibfield
  {journal} {\bibinfo  {journal} {Proc. Natl. Acad. Sci. U.S.A.}\ }\textbf
  {\bibinfo {volume} {102}},\ \bibinfo {pages} {6665} (\bibinfo {year}
  {2005})}\BibitemShut {NoStop}%
\bibitem [{\citenamefont {McGreevy}\ and\ \citenamefont
  {Pusztai}(1988)}]{McGreevy1988}%
  \BibitemOpen
  \bibfield  {author} {\bibinfo {author} {\bibfnamefont {R.~L.}\ \bibnamefont
  {McGreevy}}\ and\ \bibinfo {author} {\bibfnamefont {L.}~\bibnamefont
  {Pusztai}},\ }\href {\doibase 10.1080/08927028808080958} {\bibfield
  {journal} {\bibinfo  {journal} {Mol. Simul.}\ }\textbf {\bibinfo {volume}
  {1}},\ \bibinfo {pages} {359} (\bibinfo {year} {1988})}\BibitemShut {NoStop}%
\bibitem [{\citenamefont {Tucker}, \citenamefont {Dove},\ and\ \citenamefont
  {Keen}(2001)}]{Tucker2001}%
  \BibitemOpen
  \bibfield  {author} {\bibinfo {author} {\bibfnamefont {M.~G.}\ \bibnamefont
  {Tucker}}, \bibinfo {author} {\bibfnamefont {M.~T.}\ \bibnamefont {Dove}}, \
  and\ \bibinfo {author} {\bibfnamefont {D.~A.}\ \bibnamefont {Keen}},\ }\href
  {\doibase 10.1107/S002188980100930X} {\bibfield  {journal} {\bibinfo
  {journal} {J. Appl. Crystallogr.}\ }\textbf {\bibinfo {volume} {34}},\
  \bibinfo {pages} {630} (\bibinfo {year} {2001})}\BibitemShut {NoStop}%
\bibitem [{\citenamefont {Evrard}\ and\ \citenamefont
  {Pusztai}(2005)}]{Evrard2005}%
  \BibitemOpen
  \bibfield  {author} {\bibinfo {author} {\bibfnamefont {G.}~\bibnamefont
  {Evrard}}\ and\ \bibinfo {author} {\bibfnamefont {L.}~\bibnamefont
  {Pusztai}},\ }\href {\doibase 10.1088/0953-8984/17/5/001} {\bibfield
  {journal} {\bibinfo  {journal} {J. Phys.: Condens. Matter}\ }\textbf
  {\bibinfo {volume} {17}},\ \bibinfo {pages} {S1} (\bibinfo {year}
  {2005})}\BibitemShut {NoStop}%
\bibitem [{\citenamefont {Gereben}\ \emph {et~al.}(2007)\citenamefont
  {Gereben}, \citenamefont {J{\'{o}}v{\'{a}}ri}, \citenamefont {Temleitner},\
  and\ \citenamefont {Pusztai}}]{Gereben2007}%
  \BibitemOpen
  \bibfield  {author} {\bibinfo {author} {\bibfnamefont {O.}~\bibnamefont
  {Gereben}}, \bibinfo {author} {\bibfnamefont {P.}~\bibnamefont
  {J{\'{o}}v{\'{a}}ri}}, \bibinfo {author} {\bibfnamefont {L.}~\bibnamefont
  {Temleitner}}, \ and\ \bibinfo {author} {\bibfnamefont {L.}~\bibnamefont
  {Pusztai}},\ }\href@noop {} {\bibfield  {journal} {\bibinfo  {journal} {J.
  Optoelectron. Adv. Mater.}\ }\textbf {\bibinfo {volume} {9}},\ \bibinfo
  {pages} {3021} (\bibinfo {year} {2007})}\BibitemShut {NoStop}%
\bibitem [{\citenamefont {Opletal}\ \emph {et~al.}(2002)\citenamefont
  {Opletal}, \citenamefont {Petersen}, \citenamefont {O'Malley}, \citenamefont
  {Snook}, \citenamefont {McCulloch}, \citenamefont {Marks},\ and\
  \citenamefont {Yarovsky}}]{Opletal2002}%
  \BibitemOpen
  \bibfield  {author} {\bibinfo {author} {\bibfnamefont {G.}~\bibnamefont
  {Opletal}}, \bibinfo {author} {\bibfnamefont {T.}~\bibnamefont {Petersen}},
  \bibinfo {author} {\bibfnamefont {B.}~\bibnamefont {O'Malley}}, \bibinfo
  {author} {\bibfnamefont {I.}~\bibnamefont {Snook}}, \bibinfo {author}
  {\bibfnamefont {D.~G.}\ \bibnamefont {McCulloch}}, \bibinfo {author}
  {\bibfnamefont {N.~A.}\ \bibnamefont {Marks}}, \ and\ \bibinfo {author}
  {\bibfnamefont {I.}~\bibnamefont {Yarovsky}},\ }\href {\doibase
  10.1080/089270204000002584} {\bibfield  {journal} {\bibinfo  {journal} {Mol.
  Simul.}\ }\textbf {\bibinfo {volume} {28}},\ \bibinfo {pages} {927} (\bibinfo
  {year} {2002})}\BibitemShut {NoStop}%
\bibitem [{\citenamefont {Opletal}\ \emph {et~al.}(2008)\citenamefont
  {Opletal}, \citenamefont {Petersen}, \citenamefont {O'Malley}, \citenamefont
  {Snook}, \citenamefont {McCulloch},\ and\ \citenamefont
  {Yarovsky}}]{Opletal2008}%
  \BibitemOpen
  \bibfield  {author} {\bibinfo {author} {\bibfnamefont {G.}~\bibnamefont
  {Opletal}}, \bibinfo {author} {\bibfnamefont {T.}~\bibnamefont {Petersen}},
  \bibinfo {author} {\bibfnamefont {B.}~\bibnamefont {O'Malley}}, \bibinfo
  {author} {\bibfnamefont {I.}~\bibnamefont {Snook}}, \bibinfo {author}
  {\bibfnamefont {D.}~\bibnamefont {McCulloch}}, \ and\ \bibinfo {author}
  {\bibfnamefont {I.}~\bibnamefont {Yarovsky}},\ }\href {\doibase
  10.1016/j.cpc.2007.12.007} {\bibfield  {journal} {\bibinfo  {journal}
  {Comput. Phys. Commun.}\ }\textbf {\bibinfo {volume} {178}},\ \bibinfo
  {pages} {777} (\bibinfo {year} {2008})}\BibitemShut {NoStop}%
\bibitem [{\citenamefont {Soper}(1996)}]{Soper1996}%
  \BibitemOpen
  \bibfield  {author} {\bibinfo {author} {\bibfnamefont {A.~K.}\ \bibnamefont
  {Soper}},\ }\href {\doibase 10.1016/0301-0104(95)00357-6} {\bibfield
  {journal} {\bibinfo  {journal} {Chem. Phys.}\ }\textbf {\bibinfo {volume}
  {202}},\ \bibinfo {pages} {295} (\bibinfo {year} {1996})}\BibitemShut
  {NoStop}%
\bibitem [{\citenamefont {Soper}(2005)}]{Soper2005}%
  \BibitemOpen
  \bibfield  {author} {\bibinfo {author} {\bibfnamefont {A.~K.}\ \bibnamefont
  {Soper}},\ }\href {\doibase 10.1103/PhysRevB.72.104204} {\bibfield  {journal}
  {\bibinfo  {journal} {Phys. Rev. B}\ }\textbf {\bibinfo {volume} {72}},\
  \bibinfo {pages} {104204} (\bibinfo {year} {2005})}\BibitemShut {NoStop}%
\bibitem [{\citenamefont {Gereben}\ and\ \citenamefont
  {Pusztai}(2012)}]{Gereben2012}%
  \BibitemOpen
  \bibfield  {author} {\bibinfo {author} {\bibfnamefont {O.}~\bibnamefont
  {Gereben}}\ and\ \bibinfo {author} {\bibfnamefont {L.}~\bibnamefont
  {Pusztai}},\ }\href {\doibase 10.1002/jcc.23058} {\bibfield  {journal}
  {\bibinfo  {journal} {J. Comput. Chem.}\ }\textbf {\bibinfo {volume} {33}},\
  \bibinfo {pages} {2285} (\bibinfo {year} {2012})}\BibitemShut {NoStop}%
\bibitem [{\citenamefont {Abraham}\ \emph {et~al.}(2015)\citenamefont
  {Abraham}, \citenamefont {Murtola}, \citenamefont {Schulz}, \citenamefont
  {P{\'{a}}ll}, \citenamefont {Smith}, \citenamefont {Hess},\ and\
  \citenamefont {Lindahl}}]{GROMACS}%
  \BibitemOpen
  \bibfield  {author} {\bibinfo {author} {\bibfnamefont {M.~J.}\ \bibnamefont
  {Abraham}}, \bibinfo {author} {\bibfnamefont {T.}~\bibnamefont {Murtola}},
  \bibinfo {author} {\bibfnamefont {R.}~\bibnamefont {Schulz}}, \bibinfo
  {author} {\bibfnamefont {S.}~\bibnamefont {P{\'{a}}ll}}, \bibinfo {author}
  {\bibfnamefont {J.~C.}\ \bibnamefont {Smith}}, \bibinfo {author}
  {\bibfnamefont {B.}~\bibnamefont {Hess}}, \ and\ \bibinfo {author}
  {\bibfnamefont {E.}~\bibnamefont {Lindahl}},\ }\href {\doibase
  10.1016/j.softx.2015.06.001} {\bibfield  {journal} {\bibinfo  {journal}
  {SoftwareX}\ }\textbf {\bibinfo {volume} {1-2}},\ \bibinfo {pages} {19}
  (\bibinfo {year} {2015})}\BibitemShut {NoStop}%
\bibitem [{\citenamefont {Steinczinger}\ and\ \citenamefont
  {Pusztai}(2012)}]{Steinczinger2012}%
  \BibitemOpen
  \bibfield  {author} {\bibinfo {author} {\bibfnamefont {Z.}~\bibnamefont
  {Steinczinger}}\ and\ \bibinfo {author} {\bibfnamefont {L.}~\bibnamefont
  {Pusztai}},\ }\href {\doibase 10.5488/CMP.15.23606} {\bibfield  {journal}
  {\bibinfo  {journal} {Condens. Matter Phys.}\ }\textbf {\bibinfo {volume}
  {15}},\ \bibinfo {pages} {23606} (\bibinfo {year} {2012})}\BibitemShut
  {NoStop}%
\bibitem [{\citenamefont {Pethes}\ and\ \citenamefont
  {Pusztai}(2015)}]{Pethes2015}%
  \BibitemOpen
  \bibfield  {author} {\bibinfo {author} {\bibfnamefont {I.}~\bibnamefont
  {Pethes}}\ and\ \bibinfo {author} {\bibfnamefont {L.}~\bibnamefont
  {Pusztai}},\ }\href {\doibase 10.1016/j.molliq.2015.08.050} {\bibfield
  {journal} {\bibinfo  {journal} {J. Mol. Liq.}\ }\textbf {\bibinfo {volume}
  {212}},\ \bibinfo {pages} {111} (\bibinfo {year} {2015})}\BibitemShut
  {NoStop}%
\bibitem [{\citenamefont {Pusztai}, \citenamefont {Pizio},\ and\ \citenamefont
  {Sokolowski}(2008)}]{Pusztai2008}%
  \BibitemOpen
  \bibfield  {author} {\bibinfo {author} {\bibfnamefont {L.}~\bibnamefont
  {Pusztai}}, \bibinfo {author} {\bibfnamefont {O.}~\bibnamefont {Pizio}}, \
  and\ \bibinfo {author} {\bibfnamefont {S.}~\bibnamefont {Sokolowski}},\
  }\href {\doibase 10.1063/1.2976578} {\bibfield  {journal} {\bibinfo
  {journal} {J. Chem. Phys.}\ }\textbf {\bibinfo {volume} {129}},\ \bibinfo
  {pages} {184103} (\bibinfo {year} {2008})}\BibitemShut {NoStop}%
\bibitem [{\citenamefont {Steinczinger}\ and\ \citenamefont
  {Pusztai}(2013)}]{Steinczinger2013}%
  \BibitemOpen
  \bibfield  {author} {\bibinfo {author} {\bibfnamefont {Z.}~\bibnamefont
  {Steinczinger}}\ and\ \bibinfo {author} {\bibfnamefont {L.}~\bibnamefont
  {Pusztai}},\ }\href {\doibase 10.5488/CMP.16.43604} {\bibfield  {journal}
  {\bibinfo  {journal} {Condens. Matter Phys.}\ }\textbf {\bibinfo {volume}
  {16}},\ \bibinfo {pages} {43604} (\bibinfo {year} {2013})}\BibitemShut
  {NoStop}%
\bibitem [{\citenamefont {Soper}, \citenamefont {Bruni},\ and\ \citenamefont
  {Ricci}(1997)}]{Soper1997}%
  \BibitemOpen
  \bibfield  {author} {\bibinfo {author} {\bibfnamefont {A.~K.}\ \bibnamefont
  {Soper}}, \bibinfo {author} {\bibfnamefont {F.}~\bibnamefont {Bruni}}, \ and\
  \bibinfo {author} {\bibfnamefont {M.~A.}\ \bibnamefont {Ricci}},\ }\href
  {\doibase 10.1063/1.473030} {\bibfield  {journal} {\bibinfo  {journal} {J.
  Chem. Phys.}\ }\textbf {\bibinfo {volume} {106}},\ \bibinfo {pages} {247}
  (\bibinfo {year} {1997})}\BibitemShut {NoStop}%
\bibitem [{\citenamefont {Berendsen}, \citenamefont {Grigera},\ and\
  \citenamefont {Straatsma}(1987)}]{Berendsen1987}%
  \BibitemOpen
  \bibfield  {author} {\bibinfo {author} {\bibfnamefont {H.~J.~C.}\
  \bibnamefont {Berendsen}}, \bibinfo {author} {\bibfnamefont {J.~R.}\
  \bibnamefont {Grigera}}, \ and\ \bibinfo {author} {\bibfnamefont {T.~P.}\
  \bibnamefont {Straatsma}},\ }\href {\doibase 10.1021/j100308a038} {\bibfield
  {journal} {\bibinfo  {journal} {J. Phys. Chem.}\ }\textbf {\bibinfo {volume}
  {91}},\ \bibinfo {pages} {6269} (\bibinfo {year} {1987})}\BibitemShut
  {NoStop}%
\bibitem [{\citenamefont {Berendsen}\ \emph {et~al.}(1984)\citenamefont
  {Berendsen}, \citenamefont {Postma}, \citenamefont {van Gunsteren},
  \citenamefont {DiNola},\ and\ \citenamefont {Haak}}]{Berendsen1984}%
  \BibitemOpen
  \bibfield  {author} {\bibinfo {author} {\bibfnamefont {H.~J.~C.}\
  \bibnamefont {Berendsen}}, \bibinfo {author} {\bibfnamefont {J.~P.~M.}\
  \bibnamefont {Postma}}, \bibinfo {author} {\bibfnamefont {W.~F.}\
  \bibnamefont {van Gunsteren}}, \bibinfo {author} {\bibfnamefont
  {A.}~\bibnamefont {DiNola}}, \ and\ \bibinfo {author} {\bibfnamefont {J.~R.}\
  \bibnamefont {Haak}},\ }\href {\doibase 10.1063/1.448118} {\bibfield
  {journal} {\bibinfo  {journal} {J. Chem. Phys.}\ }\textbf {\bibinfo {volume}
  {81}},\ \bibinfo {pages} {3684} (\bibinfo {year} {1984})}\BibitemShut
  {NoStop}%
\bibitem [{\citenamefont {Wu}, \citenamefont {Tepper},\ and\ \citenamefont
  {Voth}(2006)}]{Wu2006}%
  \BibitemOpen
  \bibfield  {author} {\bibinfo {author} {\bibfnamefont {Y.}~\bibnamefont
  {Wu}}, \bibinfo {author} {\bibfnamefont {H.~L.}\ \bibnamefont {Tepper}}, \
  and\ \bibinfo {author} {\bibfnamefont {G.~A.}\ \bibnamefont {Voth}},\ }\href
  {\doibase 10.1063/1.2136877} {\bibfield  {journal} {\bibinfo  {journal} {J.
  Chem. Phys.}\ }\textbf {\bibinfo {volume} {124}},\ \bibinfo {pages} {024503}
  (\bibinfo {year} {2006})}\BibitemShut {NoStop}%
\bibitem [{\citenamefont {McGreevy}(2001)}]{McGreevy2001}%
  \BibitemOpen
  \bibfield  {author} {\bibinfo {author} {\bibfnamefont {R.~L.}\ \bibnamefont
  {McGreevy}},\ }\href {\doibase 10.1088/0953-8984/13/46/201} {\bibfield
  {journal} {\bibinfo  {journal} {J. Phys.: Condens. Matter}\ }\textbf
  {\bibinfo {volume} {13}},\ \bibinfo {pages} {R877} (\bibinfo {year}
  {2001})}\BibitemShut {NoStop}%
\bibitem [{\citenamefont {Teleman}, \citenamefont {J{\"{o}}nsson},\ and\
  \citenamefont {Engstr{\"{o}}m}(1987)}]{Teleman1987}%
  \BibitemOpen
  \bibfield  {author} {\bibinfo {author} {\bibfnamefont {O.}~\bibnamefont
  {Teleman}}, \bibinfo {author} {\bibfnamefont {B.}~\bibnamefont
  {J{\"{o}}nsson}}, \ and\ \bibinfo {author} {\bibfnamefont {S.}~\bibnamefont
  {Engstr{\"{o}}m}},\ }\href {\doibase 10.1080/00268978700100141} {\bibfield
  {journal} {\bibinfo  {journal} {Mol. Phys.}\ }\textbf {\bibinfo {volume}
  {60}},\ \bibinfo {pages} {193} (\bibinfo {year} {1987})}\BibitemShut
  {NoStop}%
\end{thebibliography}%

\end{document}